\documentclass[letterpaper,12pt]{article}
\usepackage{amssymb}

\usepackage{ifpdf}
\ifpdf
	\usepackage[pdftex]{graphicx}
	\usepackage[pdftex,unicode,implicit]{hyperref}

	\hypersetup{
  	pdftitle     = {Closure of the algebra of constraints for a nonprojectable Ho\v{r}ava model},
  	pdfkeywords  = {},
  	pdfauthor    = {J. Bellor\'{\i}n and A. Restuccia},
  	pdfcreator   = {pdf\LaTeXe\ with package \flqq hyperref\frqq},
  	pdfproducer  = {pdf\LaTeXe\ with package \flqq hyperref\frqq},
  	pdfpagemode  = UseNone,  
  	pdffitwindow = true,  
  	unicode      = true,
  	plainpages   = true,
  	colorlinks   = true,  
  	citecolor    = blue,  
  	urlcolor     = blue,
  	linkcolor    = blue
	}

\else
  \usepackage[dvips]{graphicx}
  \usepackage[unicode,implicit]{hyperref}

\fi

\makeatletter
\@addtoreset{equation}{section}
\makeatother

\begin{document}

\begin{flushright}
{\small
SB/F/382-10\\
Oct $26^{\rm th}$, $2010$}
\end{flushright}

\begin{center}

\vspace*{2cm}
{\bf \LARGE 
Closure of the algebra of constraints for a nonprojectable Ho\v{r}ava model} 
\vspace*{2cm}

{\sl\large Jorge Bellor\'{\i}n and Alvaro Restuccia}
\vspace*{1cm}

{\it Departamento de F\'{\i}sica, Universidad Sim\'on Bol\'{\i}var, Valle de Sartenejas,\\ 
1080-A Caracas, Venezuela.} \\[1ex]
{\tt jorgebellorin@usb.ve, arestu@usb.ve} \\

\vspace*{2cm}
{\bf Abstract}
\begin{quotation}{\small
We perform the Hamiltonian analysis for a nonprojectable Ho\v{r}ava model whose potential is composed of $R$ and $R^2$ terms. We show that Dirac's algorithm for the preservation of the constraints can be done in a closed way, hence the algebra of constraints for this model is consistent. The model has an extra, odd, scalar mode whose decoupling limit can be seen in a linear-order perturbative analysis on weakly varying backgrounds.

Although our results for this model point in favor of the consistency of the Ho\v{r}ava theory, the validity of the full nonprojectable theory still remains unanswered.

}\end{quotation}

\end{center}

\thispagestyle{empty}

\newpage

\section{Introduction}
Since Ho\v{r}ava formulated his theory \cite{Horava:2009uw} of a renormalizable quantum gravity based on anisotropic relativity, there has been a debate about its theoretical consistency. Several analyses have been performed on models having the anisotropic relativity symmetry based on Ho\v{r}ava's proposal, some of them can be found in Refs.~\cite{varios, Li:2009bg, Blas:2009yd, Kobakhidze:2009zr, Blas:2009qj, Bellorin:2010je}. The main discussion has been focused in the number of physical degrees of freedom, the physical consistency of their propagation, the coupling/decoupling of extra modes at the IR and the structure of the constraints together with the closure of their algebra, which is the point that  mainly concerns us in this paper\footnote{There have been also several analyses on exact classical solutions, some of them are \cite{varios:soluciones}.}.

One of the issues of the general discussion raises at the very definition of the action, since the gauge symmetry group of Ho\v{r}ava models allows to formulate them in two different ways: the projectable and nonprojectable versions. In the projectable case the Hamiltonian constraint is automatically a global constraint. As a consequence, such models have an extra scalar mode at each order in energy scale. Hence, there is not a clear way in which general relativity can be recovered in the projectable case, unless other ingredients are added to the theory \cite{Horava:2010zj}. Because of this we concentrate our study in the nonprojectable scheme.

In a previous paper \cite{Bellorin:2010je} we completed successfully the Hamiltonian analysis for the lowest-order effective action of the Ho\v{r}ava theory. This action, which we call the $\lambda R$-model, is given by
\begin{equation}
S = 
\int dt d^3 x \sqrt{g} N ( K_{ij} K^{ij} - \lambda K^2 + R ) 
\label{raction}
\end{equation}
and it is the action one obtains if the potential of the original Ho\v{r}ava theory is truncated up to second order in spatial derivatives\footnote{The Hamiltonian analysis for the projectable counterpart of the action (\ref{raction}) was done in Ref.~\cite{Kobakhidze:2009zr}.}. This action differs from the Arnowitt-Deser-Misner (ADM) formulation of Einstein's general relativity unless $\lambda = 1$. Indeed, the presence of the constant $\lambda$ breaks the gauge symmetry of general space-time diffeomorphisms and due to this it was believed that the effective theory (\ref{raction}) reproduces general relativity only when $\lambda = 1$. However, our analysis \cite{Bellorin:2010je} showed that this is not the case; the dynamics of the action (\ref{raction}) is identical to the one of general relativity, regardless of the value of $\lambda$.

Let us take again and briefly the main point of the approach of Ref.~\cite{Bellorin:2010je}. When the preservation in time of the Hamiltonian constraint, which becomes a second-class constraint due to the reduced gauge symmetry group of Ho\v{r}ava models, is imposed, a new constraint emerges,
\begin{equation}
 \nabla_i ( N^2 \nabla^i \pi ) =  0 \,.
 \label{cr}
\end{equation}
Throughout this paper we shall call constraint $\mathcal{C}$ to the condition emerging when the preservation of the Hamiltonian constraint is imposed. Here $\pi$ is the trace of the momentum conjugated to $g_{ij}$. From Eq.~(\ref{cr}) and the asymptotic behavior of the field variables it is straightforward to derive a global condition, which is
\begin{equation}
\int \frac{d^3 x}{\sqrt{g}} {N^2 \left(\nabla_i \pi \right)^2}
= 0 \,.
\end{equation}
This equation leads unavoidably to the condition $\pi = 0$ at each point of space-time. In this point we would like to remark that the possibility of $N = 0$ is discarded at the very definition of the action (\ref{raction}), whose kinetic term depends on $N$ as $\sim 1/N$. The same restriction must be imposed on the Hamiltonian, otherwise it lacks its relation to the action (\ref{raction}). Another important consideration is that this result is independent of the role one assigns to the variable $N$ in the canonical analysis: the result is the same if $N$ is taken as a canonical variable or as a Lagrange multiplier, as we are going to see in Sec. II. Therefore, constraint $\mathcal{C}$ becomes completely equivalent to $\pi = 0$ in the $\lambda R$-model, such that the kinetic trace-term drops out from the action (\ref{raction}). We stress that $\pi = 0$ is not an additional constraint put by hand. Its preservation in time yields an elliptic partial differential equation (PDE) for $N$, such that Dirac's algorithm stops here.

When one focuses the problem from the side of general relativity, it turns out that $\pi = 0$ is a partial gauge fixing condition that can be imposed to fix the part of the gauge symmetry that mixes time with space and that is absent in the Ho\v{r}ava theory. When this partial gauge fixing condition is imposed on the side of general relativity, it becomes clear the equivalence between it and the $\lambda R$-model.

Naturally, one question arises: can the same approach of the $\lambda R$-model be extended to other models with higher-derivative potentials?. The aim of this paper
is to give an answer to this question by performing completely Dirac's algorithm of constraints for a model with a potential that goes one step beyond the previous $\lambda R$-model: we shall consider a potential with a $R^2$ term, which we shall call the $\lambda R^2$-model. Thus, we shall study a fourth-order potential having a mathematical structure that is closer to the full Ho\v{r}ava theory than the $\lambda R$-model.

We give an advance of our main results: Dirac's program can be performed successfully for the $\lambda R^2$-model, showing in this way the closure of the algebra. Constraint $\mathcal{C}$ is again solved completely as an elliptic PDE for $\pi$, but now the solutions depend in general on $N$. This feature departs from the $\lambda R$-model and leads to the presence of one extra scalar degree, which is odd in the number of initial data it propagates. We also show a perturbative scheme under which this extra mode decouples at low energies.

In Ref.~\cite{Blas:2009yd} an extra, odd scalar mode was also found for a model with a $R_{ij} R^{ij}$ term in the potential. By making a perturbative analysis, those authors argued that the extra mode makes the original Ho\v{r}ava theory inconsistent. In particular, they concluded that the extra mode presents very fast instabilities and strong coupling at the IR. However, from their analysis one cannot arrive at that conclusion since their specific studies about the propagation of the extra mode were restricted to the $\lambda R$-model. Indeed, the main analysis of Ref.~\cite{Blas:2009yd} concerning the perturbative propagation of the extra mode was based on the St\"uckelberg formalism, which allows to cast the original Ho\v{r}ava model as a totally covariant theory using an extra scalar field. For the perturbative analysis those authors only took the reduced action given in Eq.~(32) of their paper, which does not consider the dynamics of the graviton because the 4D scalar curvature term is not taken into account. When one adds the 4D scalar curvature term and imposes the unitary gauge it turns out that the resulting action is exactly the $\lambda R$-model. Therefore, the perturbative analysis done in Ref.~\cite{Blas:2009yd} cannot yield any information about the extra mode since it was performed at the order of the $\lambda R$-model and this theory does not possesses it\footnote{In addition, in Ref.~\cite{Blas:2009yd} a preliminary perturbative analysis was done besides the St\"uckelberg approach. There those authors presented a dispersion relation for the extra mode. However, that study was again restricted to the $\lambda R$-model, as they indicated in their paper. Indeed, If one takes Eqs.~(21) and (22) of Ref.~\cite{Blas:2009yd} and for the background imposes $\bar{K} = 0$, which is the only solution of the $\mathcal{C}$ constraint of the $\lambda R$-model, one finds from Eq.~(22) that $\kappa = 0$ and from Eq.~(21) $n = 0$, $\kappa$ being the perturbation of $\bar{K}$ and $n$ the one of $\bar{N}$. This result is consistent with the fact that the $\lambda R$-model has no an extra mode. Hence, the dispersion relation (23) and all the analysis that follows is invalid.}. Of course, at the moment those authors presented their study it seemed a consistent and convenient approach to analyze the extra mode using only the effective $\lambda R$-model since the presence of the constant $\lambda$ apparently entails a deviation from general relativity. After \cite{Bellorin:2010je} we know that there is not an extra mode nor a deviation from general relativity for any value of $\lambda$.

This paper is organized as follows: in Sec.~II we perform completely Dirac's algorithm for the $\lambda R^2$-model. In Sec.~III we carry out a perturbative analysis for the constraint $\mathcal{C}$ and its preservation in time. To achieve this we take the approach of the weakly varying background used in Ref.~\cite{Blas:2009yd}, showing the decoupling of the extra mode at linear oder in perturbation and zeroth- and first-order in derivatives of the background. In Sec.~IV we present our conclusions.


\section{Analysis of the constraints}
The action of the $\lambda R^2$-model, in ADM variables, is
\begin{equation}
S = 
\int dt d^3 x \sqrt{g} N ( G^{ijkl} K_{ij} K_{kl}  + R + \zeta R^2) \,,
\label{r2action}
\end{equation}
where $R$ is the spatial Ricci scalar, $\zeta$ is a coupling constant and the kinetic variable $K_{ij}$ is the extrinsic curvature defined by
\begin{equation}
 K_{ij} = 
 \frac{1}{2N} ( \dot{g}_{ij} - 2 \nabla_{(i} N_{j)} ) \,.
\label{k}
\end{equation}
Of course, this requires that $ N\neq 0$ in all space-time. $G^{ijkl}$ is a four-indices metric,
\begin{equation}
G^{ijkl} = 
\frac{1}{2} (g^{ik} g^{jl} + g^{il} g^{jk} )
- \lambda g^{ij} g^{kl} \,.
\end{equation}
For the case of $\lambda \neq {1}/{3}$ the inverse of $G^{ijkl}$ is given by
\begin{equation}
\mathcal{G}_{ijkl} = 
\frac{1}{2} (g_{ik} g_{jl} + g_{il} g_{jk} )
- \frac{\lambda}{3\lambda - 1} g_{ij} g_{kl} \,.
\label{inverseg}
\end{equation}

The momenta conjugated to $N$ and $N_i$ vanish automatically for the Lagrangian given in Eq.~(\ref{r2action}). The same happens in general relativity and in any Ho\v{r}ava model. Because of this $N$ and $N_i$ are usually considered as Lagrange multipliers, as we did for the $\lambda R$-model in Ref.~\cite{Bellorin:2010je}. However, for the $\lambda R^2$-model we find it convenient to keep the lapse function $N$, together with its canonical conjugated, as part of the canonical variables, whereas the shift functions $N_i$ are regarded as Lagrange multipliers and their conjugated momenta are dropped out from the canonical space. We denote by $\phi$ the momentum conjugated to $N$ and by $\pi^{ij}$ the one conjugated to $g_{ij}$, which is found to be
\begin{equation}
\frac{\pi^{ij}}{\sqrt{g}} = 
G^{ijkl} K_{kl} \,.
\label{pi}
\end{equation}
This formula is the same for any theory having the same kinetic term of the Ho\v{r}ava theory. The velocities $\dot{g}_{ij}$ can be solved explicitly in terms of the momentum $\pi^{ij}$ whenever $\lambda \neq 1/3$.

We assume that the whole spatial manifold is noncompact. All the configurations are fixed at infinity and correspond to Minkowski space-time. We also assume that the field variables have the same asymptotic behavior of general relativity: in asymptotically flat coordinates, as $r$ approaches to infinity, they behave as
\begin{equation}
\begin{array}{rclrcl}
 g_{ij} & = & \delta_{ij} + \mathcal{O}(r^{-1}) \,, 
\hspace{2em}
 & \pi^{ij} & = & \mathcal{O}(r^{-2}) \,,
\\[2ex]
 N & = & 1 + \mathcal{O}(r^{-1}) \,, 
 & N_i & = & \mathcal{O}(r^{-1}) \,.
\end{array}
\label{asymptotics}
\end{equation}
We need to clarify that here the notation $\mathcal{O}(r^{-\alpha})$ means that the variable asymptotically receives contributions at most of order $r^{-\alpha}$, but they can be effectively lower. For example, in the $\lambda R$-model if $\pi|_{\infty} = 0$ then it is zero everywhere, hence it does not receive any contribution.

By performing the Legendre transformation, assuming the $\lambda \neq 1/3$ condition, we obtain the Hamiltonian
\begin{equation}
\begin{array}{rcl}
H & = &
\left< N \mathcal{H} + N_i \mathcal{H}^i + \sigma \phi \right> \,,
\\[2ex]
\mathcal{H} & \equiv &
\mathcal{G}_{ijkl} {\displaystyle\frac{\pi^{ij} \pi^{kl}}{\sqrt{g}}} 
 - \sqrt{g} ( R + \zeta R^2 ) \,,
\\[2ex]
\mathcal{H}^i & \equiv &
- 2 \nabla_j \pi^{ji} \,,
\end{array}
\label{hamiltonian}
\end{equation}
where we use the brackets $\left<{\:}\right>$ to denote integration over all the spatial manifold, $\left< \mathcal{F} \right> \equiv \int \mathcal{F} d^3 x$, and the standard definition for covariant derivatives on densities is understood. We have added up the primary constraint $\phi$, $\sigma$ being its Lagrange multiplier, such that the above is the total Hamiltonian. Under spatial coordinate transformations $\phi$ behaves as a scalar-density, hence $\sigma$ is a scalar. Notice that this Hamiltonian has been deduced under the assumption $N \neq 0$, thus we must keep this restriction in the Hamiltonian in order to be able to associate it to the $\lambda R^2$-model.

Now we start the Dirac's algorithm to extract all the constraints of the theory and to ensure their preservation in time. The condition $\mathcal{H}^i = 0$ emerges as a primary constraint when one takes variations of the canonical action w.r.t.~$N_i$. The preservation in time of this constraint is ensured due to its role as the generator of the spatial coordinate transformations \cite{DeWitt:1967yk}. Hence, as the next step we must demand the preservation in time of the primary constraint $\phi$, which yields
\begin{equation}
 \{ \phi , H \} = - \mathcal{H} \,.
\end{equation}  
Therefore $\mathcal{H}$ is a secondary constraint of the theory.

The preservation in time of $\mathcal{H}$ requires to compute only the bracket of $\mathcal{H}$ with itself, $\{ \mathcal{H} , \left< N \mathcal{H} \right> \}$. We obtain that the vanishing of this bracket leads to the secondary constraint $\mathcal{C} = 0$, where
\begin{eqnarray}
 \mathcal{C} & \equiv & \nabla_i ( N^2 \gamma^i )  \,,
\label{c}
 \\[1ex]
 \gamma^i & \equiv & 
 \beta \rho^2 \nabla^i ( \pi / \rho ) + 2 \zeta \pi^{ij} \nabla_j R 
\label{gamma2}
\end{eqnarray}
and in addition $\pi \equiv g_{ij} \pi^{ij}$, $\rho \equiv 1 + 2 \zeta R$ and $\beta \equiv (\lambda -1)/(3\lambda -1)$. Notice that expression (\ref{gamma2}) is regular in the limit $\rho = 0$. Constraint $\mathcal{C}$ is the extension of the extra constraint found in the $\lambda R$-model \cite{Blas:2009yd,Bellorin:2010je}. It arises as a consequence of the lacking of the general space-time diffeomorphisms as gauge symmetries, which leads to the second-class behavior of the Hamiltonian constraint $\mathcal{H}$. 

Having found the constraint $\mathcal{C}$, in order to obtain a consistent closure of the algebra it becomes crucial to determine the appropriate variable for which its general solution can be found in a closed way. We recall that we have adopted a scheme under which $N$, as well as $g_{ij}$ and $\pi^{ij}$, are canonical variables, hence all of them are treated on the same footing. This allows us to solve, in principle, any of these variables in terms of the other ones.

The first thing to do is contrasting with the $\lambda R$-model: if the vector-density field $\gamma^i$ has the form $\gamma^i = \chi^2 \nabla^i \varphi$ for some nonzero function $\chi$ and a function $\varphi$ vanishing at infinity, then, by following the same steps done in Ref.~\cite{Bellorin:2010je}, one easily obtains that the constraint $\mathcal{C}$ does not give any information about $N$, it instead implies $\varphi = 0$. This is the case for the $\lambda R$-model, where $\gamma^i = \beta \nabla^i \pi$. For the $\lambda R^2$-model, the first term in $\gamma^i$ (\ref{gamma2}) has exactly the required form, but the last term does not, hence we are faced with an obstruction to apply this procedure. This leads us to conclude that there are solutions of $\mathcal{C}$ with nonvanishing vector-density field $\gamma^i$ which are $N$-dependent.

The next question we might ask is if the general solution of $\mathcal{C}$ can be cast as a condition for $N$. We are going to show that the answer is negative by making use of the asymptotic expansion of the constraint $\mathcal{C}$ for large $r$. This is based on the fact that the values of the field variables at infinity are fixed and correspond exactly to Minkowski space-time. Thus, the degrees of freedom of any field variable are expressed asymptotically in the coefficient of its first corrections. By using the expansion (\ref{asymptotics}) in (\ref{gamma2}) we obtain that $\gamma^i = \mathcal{O}(r^{-3})$ asymptotically. Then, it is clear that the leading term in the asymptotic expansion of $\mathcal{C}$, which is of order $\mathcal{O}(r^{-4})$, comes from the pure divergence of $\gamma^i$ and does not involve the variable $N$. Consequently, the general solution of constraint $\mathcal{C}$ necessarily restricts $\gamma^i$, hence $g_{ij}$ and $\pi^{ij}$, at least asymptotically.

In order to solve constraint $\mathcal{C}$ in the most general way, we cast it as an equation for $\pi$. It results that for this variable $\mathcal{C}$ is an elliptic PDE compatible with the asymptotic conditions. To see this it is convenient to make the covariant decomposition of $\pi^{ij}$ between its trace and traceless parts,
\begin{equation}
 \pi^{ij} = 
\pi^{ij}_{\mathrm{T}} + {\textstyle\frac{1}{3}} g^{ij} \pi \,.
\end{equation}
Using this decomposition, the constraint $\mathcal{C}$ given in Eqs.~(\ref{c}) and (\ref{gamma2}) takes the form
\begin{equation}
 \left[ \beta N^2 \rho^2 \nabla^2
+ \left( \beta \partial^i (N^2 \rho^2)
  + {\textstyle\frac{1}{6}} N^2 \partial^i \rho^2 \right) \nabla_i
+ {\textstyle\frac{1}{6}} \nabla_i ( N^2 \partial^i \rho^2 ) \right] (\pi / \rho) =
 - 2 \zeta \nabla_i (N^2 \pi^{ij}_{\mathrm{T}} \partial_j R ) \,,
\end{equation}
which, for $\lambda \neq 1$, is an elliptic PDE for $\pi / \rho$ with an $N$-dependent source term. There always exists a unique solution for it satisfying the boundary conditions. The interpretation of $\mathcal{C}$ we make here is similar to its role in the $\lambda R$-model, with the difference that here the solution for $\pi$ depends on $N$. This is the main reason why it is convenient to regard $N$ as a canonical variable in the $\lambda R^2$-model.

Now we impose the preservation in time of the constraint $\mathcal{C}$ (\ref{c}), which is established by the condition $\{ \mathcal{C} , \left< N\mathcal{H} + \sigma \phi \right> \} = 0$. We evaluate this condition by direct computations and simplifying the result using all the known constraints. This yields an equation for $\sigma$:
\begin{equation}
 V^i \nabla_i (\sigma / N ) = - {\textstyle\frac{1}{2}}
 \nabla_i \left[ N^2 (\beta \rho^2 \mathcal{A}^i + \zeta \mathcal{B}^i 
 + 4 \zeta g^{-1} \nabla_j \mathcal{D}^{ij} ) \right] \,,
\label{sigmaeq}
\end{equation}
where
\begin{equation}
\begin{array}{rcl}
V^i & \equiv & N^2 \gamma^i / \sqrt{g} \,,
\\[2ex]
\mathcal{A}^i & \equiv &
  -2 \nabla^i \left[ \nabla^2 (N\rho) - N R / \rho \right]
  + {\displaystyle \frac{2 N \pi}{(3\lambda - 1) g}   \nabla^i (\pi/\rho)
  + \frac{\pi/\rho}{(3\lambda - 1)g}  \nabla^i (N\pi) \,,}
\\[2ex]
  \mathcal{B}^i & \equiv &
  \left[ N( R + \zeta R^2 ) - \nabla^2 (N\rho) \right] \nabla^i R
  + 2 \left[ \nabla^i \nabla^j (N\rho) - N \rho R^{ij} \right] \nabla_j R
\\[2ex]
& &
  + \nabla^i [ N R (R + \zeta R^2) ]
  + \beta \rho^2 \nabla^i ( N R^2 / \rho )
\\[2ex]
& & + {\displaystyle\frac{4 \beta \rho}{g} \left[ 
   2 \left( \pi^{kl} \nabla_k \nabla_l N 
   - \beta \nabla^2 (N\pi) 
   - N \mathcal{G}_{klmn} \pi^{kl} R^{mn}\right) 
  \nabla^i (\pi/\rho) \right.}
\\[2ex]
& & \left.
   + \rho \nabla^i ( N\pi \mathcal{G}_{jklm} \pi^{jk} R^{lm} /\rho^2)
   - \rho \nabla^i (\pi \pi^{kl} \nabla_k \nabla_l N / \rho^2)
   + \beta \rho \nabla^i \left( \pi \nabla^2 (N\pi) / \rho^2 \right) \right] \,,
\\[2ex]
\mathcal{D}^{ij} & \equiv &
      \pi^{ij} \left( N \pi^{kl} R_{kl} 
    - \pi^{kl} \nabla_k \nabla_l N
    + \beta \nabla^2 (N\pi) \right) 
    - N R \pi^{k(i} \pi^{j)}{}_{k} \,.
 \end{array}
\end{equation}

Equation (\ref{sigmaeq}) defines a first-order PDE for $\sigma$ in any open set in which the vector field $V^i$ is nonvanishing. The standard theorems about PDEs ensure the existence and uniqueness of the solution once a suitable boundary condition has been provided. Since $V^i \nabla_i$ is a flow-operator, the boundary data compatible with it consists of giving the value of $\sigma / N$ on a noncharacteristic 2D surface, that is, a surface that is nonparallel to $V^i$ at each point. Notice that, due to constraint $\mathcal{C}$, the vector field $V^i$ is divergenceless, and by the required boundary conditions it vanishes at infinity with a decay of order $\mathcal{O}(r^{-3})$. As a consequence, it can be shown that its flow cannot reach infinity. Further details about the flow of $V^i$ and the way in which the Eq.~(\ref{sigmaeq}) must be solved can be found in the Appendix.

We wish to clarify the compatibility between the generic solutions of Eq.~(\ref{sigmaeq}) and the expected asymptotic behavior of the field variables. To this end we first point out that the canonical equation of motion $\delta S / \delta \phi$ yields
\begin{equation}	
 \dot{N} = \sigma \,,
\label{dotn}
\end{equation}
which gives the evolution of the variable $N$ once $\sigma$ has been solved from the Eq.~(\ref{sigmaeq}). From Eqs.~(\ref{asymptotics}) and (\ref{dotn}) we deduce that asymptotically $\sigma$ behave as $\sigma = \mathcal{O}(r^{-1})$, such that the l.h.s.~of Eq.~(\ref{sigmaeq}), which involves $\sigma$, has an asymptotic decay of order $\mathcal{O}(r^{-5})$. In the r.h.s., the leading term comes from the first term of $\mathcal{A}^i$, which also leads to a decay of order $\mathcal{O}(r^{-5})$ in the r.h.s.~of Eq.~(\ref{sigmaeq}). Thus we see that the Eq.~(\ref{sigmaeq}) imposes consistent restrictions on the multiplier $\sigma$ at any order asymptotically. We conclude that Eq.~(\ref{sigmaeq}) can be consistently used to solve for the Lagrange multiplier $\sigma$, ending Dirac's algorithm with this step.

We have then ended up with the first-class constraint $\mathcal{H}^i = 0$ (3 constraints) and with the second-class ones $\phi = 0$, $\mathcal{H} = 0$ and $\mathcal{C} = 0$ (3 constraints). The canonical variables are $\{ N, \phi, g_{ij}, \pi^{ij}\}$ (14 variables). The number of physical degrees of freedom in the canonical space is given by
\begin{equation}
\begin{array}{l}
\mbox{(\# Physical D.O.F.)} =  
\nonumber \\[1ex]
 \hspace*{2em}  \mbox{(\# Canonical var.)} 
   - 2 \times \mbox{(\# 1$^{\mbox{\tiny st}}$ class const.)} 
     -\mbox{(\# 2$^{\mbox{\tiny nd}}$ class const.)} 
     = 5 \,.
\end{array}
\end{equation}
Four of these 5 degrees of freedom correspond to the propagation of the graviton in the phase space. The remaining degree is an odd scalar mode, that is, a scalar field that propagates only one Cauchy field as initial data, but not its time-derivative. This extra mode is represented by the canonical variable $N$ and the corresponding evolution equation is given in (\ref{dotn}). The general analysis about the propagation of the extra mode is as follows: once $\sigma$ has been solved from Eq.~(\ref{sigmaeq}) on a $t = \mbox{constant}$ slice $\mathcal{M}_3$, then the evolution of $N$ is determined by Eq.~(\ref{dotn}) together with the initial data $N(t=0,\vec{x})$ provided. This initial data represents extra information in the sense that it does not arise in general relativity.

It is worth mentioning that our analysis is based on the fact that the solution for the multiplier $\sigma$ is constructed by following the integral lines of the vector field $V^i$. There could be, of course, some configurations for which $V^i$ vanishes in all points of space-time. Those points of the phase space are very special since they solve automatically the constraint $\mathcal{C}$ for arbitrary $N$ and also the Eq.~(\ref{sigmaeq}) cannot be used to solve for $\sigma$ in any open set. As an example, this is the case of all the configurations satisfying $ R = -1/2\zeta$ ($\rho =  0$) at each point. The algebra of constraints on those configurations deserves a detailed analysis besides the study we present here.

As a check of consistency we now go to the limit $\zeta = 0$ and compare with the $\lambda R$-model. In this limit the Hamiltonian constraint of the $\lambda R$-model is recovered from (\ref{hamiltonian}) and $\gamma^i$ given in Eq.~(\ref{gamma2}) is reduced to $\gamma^i = \beta \nabla^i \pi$, such that $\mathcal{C}$ also takes the same form of the $\lambda R$-model and $\pi = 0$ everywhere is its only solution satisfying (\ref{asymptotics}) asymptotically. However, note that in the scheme we have adopted in this paper $N$ is a canonical variable, as a consequence Eq.~(\ref{sigmaeq}) must still be analyzed. This equation simplifies greatly since $\gamma^i$ vanishes in the $\zeta = 0$ limit due to the $\mathcal{C}$ constraint. The l.h.s.~of Eq.~(\ref{sigmaeq}) vanishes, consequently we do not get from it an equation for $\sigma$, but an extra constraint,
\begin{equation}
 \beta \nabla_i \left[ N^2 \nabla^i \left( \nabla^2 - R \right) N \right]
 = 0 \,.
\label{nequation}
\end{equation}
This constraint is solved in the same way one solves $\mathcal{C}$ in the $\lambda R$-model: the scalar $(\nabla^2  - R) N$ is of order $\mathcal{O}(r^{-3})$ asymptotically; one  multiplies Eq.~(\ref{nequation}) by it and integrates over all the spatial manifold. From the resulting integral, after an integration by parts, one deduces that the only solution is (for $\lambda \neq 1$)
\begin{equation}
 (\nabla^2  - R) N = 0 
\label{nequation2}
\end{equation}
at each space-time point, which is exactly the equation that determines $N$ in the $\lambda R$-model \cite{Bellorin:2010je}. In that reference this equation was obtained by requiring the preservation in time of the constraint $\pi = 0$. In the $\zeta = 0$ limit of the $\lambda R^2$-model the mechanism works analogously, but being careful in identifying the Eq.~(\ref{sigmaeq}) as an equation for $N$, while $\sigma$ decouples completely from it. Equation (\ref{nequation2}) must be regarded as a further extra constraint of the theory, such that the number of degrees of freedom is reduced in one. In this scheme the Lagrange multiplier $\sigma$ will be determined once the preservation in time of the constraint (\ref{nequation2}) is imposed (or directly by Eq.~(\ref{dotn}), which yields the same result).


\section{Perturbative analysis on a weakly varying background}
The general analysis done in the previous section has revealed the presence of an extra, odd, scalar mode. However, this extra mode is not found when one analyses the dynamics of the lowest-order effective theory, the $\lambda R$-model, taken as an autonomous theory \cite{Bellorin:2010je}. Therefore, it becomes crucial to determine if the extra mode we have found in the $\lambda R^2$-model decouples at low energies, such that the physics of the $\lambda R$-model is smoothly recovered in the IR limit.

A perturbative analysis is a good approach to elucidate this point. The conditions that play a role in the coupling/decoupling of the extra mode are the constraint $\mathcal{C}$ and its preservation in time, which yields the Eq.~(\ref{sigmaeq}). We may concentrate ourselves only in the perturbative expansions of these conditions and in particular we must study the perturbative behavior of the variables $N$ and $\sigma$. With regard to the background, our main interest is not in the specific evolution of the modes around a given vacuum, we are more interested in seeing the IR decoupling of the extra mode for a broad class of backgrounds. To achieve this, we adopt the scheme used in Ref.~\cite{Blas:2009yd}: instead of dealing with a specific background, we assume that a solution of the constraints and the equations of motion, denoted by $\{\bar{g}_{ij}, \bar{\pi}^{ij}, \bar{N}, \bar{\sigma}\}$, exists and expand the fields around the solution up to linear order,
\begin{equation}
\begin{array}{rclrcl}
 g_{ij} & = & \bar{g}_{ij} + h_{ij} \,,
 \hspace{2em}
 & \pi^{ij} & = & \bar{\pi}^{ij} + p^{ij} \,,
 \\[1.5ex]
 N & = & \bar{N} + n \,,
 & 	\sigma & = & \bar{\sigma} + \tau \,.
 \end{array}
\label{perturbation}
\end{equation}
It is supposed that the solution depends both on space and time, hence all of the fields of the background are in principle different from zero. In order to further simplify the analysis, it is also assumed that the background is a \emph{weakly varying configuration}, in particular when it is compared with the space-time scales in which the perturbations varies. Specifically, the background metric, together with $\bar{N}$, changes at a typical space-time scale $L$. This gives $\bar{\pi}^{ij} \sim L^{-1}$, $\bar{R} \sim L^{-2}$, $\partial_i \bar{N} \sim L^{-1}$ and $\bar{\sigma} \sim L^{-1}$. Under this assumption one can assign a weight to each term in the perturbative expansions of the constraints/equations of motion according to its order in $L^{-1}$.

We perform a linear-order perturbation on the constraint $\mathcal{C}$ given in Eqs.~(\ref{c}) and (\ref{gamma2}) and obtain
\begin{equation}
\mathcal{C} =
 \partial_i ( \bar{N}^2 \gamma^i ) 
 + 2\partial_i\left( \bar{N} \bar{\gamma}^i n \right) \,,
\label{cexpanded}
\end{equation}
where
\begin{equation}
\begin{array}{rcl}
\gamma^i & = &
  \gamma^{i(0)} + \gamma^{i(1)} + \gamma^{i(2)} + \gamma^{i(3)} + \gamma^{i(4)} \,,
\\[2ex]
\gamma^{i(0)} & = &
  \beta \bar{\nabla}^i p \,,
\\[2ex]
\gamma^{i(1)} & = &
    \beta \bar{\pi}^{kl} \bar{\nabla}^i h_{kl} 
  - 2 \beta \zeta \bar{\pi} \partial^i \delta R
  + 2\zeta \bar{\pi}^{ij} \partial_j \delta R
  - \beta \bar{\pi} \bar{g}^{ij} \delta\Gamma_{jk}{}^k \,,
\\[2ex]
\gamma^{i(2)} & = &
    2\beta \zeta \bar{R} \bar{\nabla}^i p 
  + 2\beta \zeta  \bar{\nabla}^i \bar{\pi} \delta R
  - \beta \bar{\nabla}_j \bar{\pi} h^{ij}
  + \beta \bar{\nabla}^i \bar{\pi}^{kl} h_{kl} \,,
\\[2ex]
\gamma^{i(3)} & = &
 2\zeta \left(
    \beta \bar{R} \bar{\pi}^{kl} \bar{\nabla}^i h_{kl}
  - \beta \partial^i \bar{R} p 
  + p^{ij} \partial_j \bar{R}
  - \beta \bar{R} \bar{\pi} \bar{g}^{ij} \delta \Gamma_{jk}{}^k \right) \,,
\\[2ex]
\gamma^{i(4)} & = &  
 2\zeta \left( 
    \beta \bar{R} \bar{\nabla}^i \bar{\pi}^{kl} h_{kl}
  - \beta \partial^i \bar{R} \bar{\pi}^{kl} h_{kl}
  - \beta \bar{R} \bar{\nabla}_j \bar{\pi} h^{ij}
  + \beta \bar{\pi} \partial_j \bar{R} h^{ij} \right) \,,
\\[2ex]
\delta \Gamma_{ij}{}^k & = &
    \bar{\nabla}_{(i} h_{j)}{}^k
  - \frac{1}{2} \bar{\nabla}^k h_{ij} \,,
\\[2ex]
\delta R & = &
 \bar{\nabla}_i \bar{\nabla}_j h^{ij} - \bar{\nabla}^2 h - \bar{R}_{ij} h^{ij} \,.
 \end{array}
 \label{gammaexpanded}
\end{equation}
In these expressions all spatial indices are raised/lowered with $\bar{g}_{ij}$, $h \equiv \bar{g}^{ij} h_{ij}$ and $p \equiv \bar{g}_{ij} p^{ij}$. In (\ref{gammaexpanded}) we have grouped\footnote{Actually, the classification is a bit disordered since covariant derivatives of the perturbative variables are not of homogeneous order in $L^{-1}$: terms having the background connection are of one order higher than the pure derivative term. This only means that in the expansion of $\gamma^i$ these terms do not belong to the order they have been written, but to the next order.} the terms of $\gamma^i$ according to their order in $L^{-1}$ and all the terms that are of order zero in the perturbative variables have been omitted because the background is a solution of the constraints of the theory.

In the expansion (\ref{cexpanded}) the perturbative variable $n$ arises only in the second term. Thus, the term of lowest order in $L^{-1}$ containing $n$ is $ + 2\beta \bar{N} \bar{\nabla}^i \bar{\pi} \partial_i n$, which is of order $L^{-2}$. Therefore, in the linear-order perturbative analysis the variable $n$ decouples from the constraint $\mathcal{C}$ if we expand it up to order $L^{-1}$, resulting in a constraint for the variables $h_{ij}$ and $p^{ij}$,
\begin{equation}
 \mathcal{C}  = \mathcal{C}^{(0)} + \mathcal{C}^{(1)} \,,
\end{equation}
where
\begin{eqnarray}
\mathcal{C}^{(0)} & = & 
 \beta \bar{N}^2 \partial_i \partial^i p \,,
\\[1ex]
\mathcal{C}^{(1)} & = &
\beta \bar{N}^2 \partial_i \ln ( \bar{N}^2 / \sqrt{\bar{g}} ) \partial^i p 
\nonumber \\[1ex] & &
 + \bar{N}^2 \left( \beta \bar{\pi}^{kl} \partial_i \partial^i h_{kl} 
 - 2 \beta \zeta \bar{\pi} \partial_i \partial^i \delta R 
 + 2 \zeta \bar{\pi}^{ij} \partial_i \partial_j \delta R 
 - \beta \bar{\pi} \partial^i \delta \Gamma_{ij}{}^j \right) \,.
\end{eqnarray}
We notice that, when expanded up to order zero in $L^{-1}$, constraint $\mathcal{C}$ acquires exactly the same perturbative form of the $\lambda R$-model: it becomes $\mathcal{C}^{(0)} = 0$ and its only solution vanishing at infinity is $p = 0$ everywhere.

Now we move to the Eq.~(\ref{sigmaeq}). For our purposes it is enough to expand  this equation up to linear order in $L^{-1}$. This yields\footnote{The expansion has been computed assuming $\lambda \neq 1$. There is not any contribution to the Eq.~(\ref{sigmaeq}) at the orders we are considering if $\lambda = 1$.}
\begin{equation}
\begin{array}{r}
 \partial^i \partial_i \bar{\nabla}^2 n 
 + \partial_i \ln (\sqrt{\bar{g}} \bar{N}^2 ) \partial^i \partial^j \partial_j n 
  + 2\zeta \bar{N} \partial^i \partial_i \bar{\nabla}^2 \delta R
  + 2\zeta ( 6 \partial_i \bar{N} + \bar{N} \partial_i \ln \sqrt{\bar{g}} )
       \partial^i \partial^j \partial_j \delta R
\\[2ex]
  - \bar{N} \partial^i \partial_i \delta R
  - ( 4 \partial^i \bar{N} + \bar{N} \partial_i \ln\sqrt{\bar{g}} )
       \partial_i \delta R 
  - \partial_m \bar{N} \bar{g}^{kl} \partial_i \partial^i \delta \Gamma_{kl}{}^m
 \hspace*{8em}
\\[2ex]
 - {\displaystyle \frac{2 \zeta \bar{N}}{\bar{g}} \left[
    \beta \bar{\pi} (\partial^i \partial_i)^2 p
    + \bar{\pi}^{ij} \partial_i \partial_j \partial^k \partial_k p \right]
 - \left[\frac{3\bar{N}\bar{\pi}}{2(3\lambda - 1)\bar{g}} 
     + \frac{\bar{\sigma}}{\sqrt{\bar{g}} \bar{N}} \right] 
      \partial^i \partial_i p
  \; = \; 0} \,.
\label{neq}
\end{array}
\end{equation}
As we may see, the perturbative multiplier $\tau$ does not arise in this equation, which is a fourth-order elliptic PDE for $n$. This result is consistent with the decoupling of the extra mode from the constraint $\mathcal{C}$. Equation (\ref{neq}) reduces in one the number of physical degrees in the phase space, showing the consistent decoupling of the extra mode at linear order in perturbation theory and zeroth and linear order in $L^{-1}$.

Let us discuss the smooth decoupling of the extra mode from the point of view of the initial Cauchy data. As we pointed out in previous section, the initial data $N(t=0,\vec{x})$ is needed in the general analysis for the complete evolution of the system. However, in perturbation theory the Lagrange multiplier $\tau$ is fixed by Eq.~(\ref{sigmaeq}) only at second order in $L^{-1}$ and higher, whereas at zeroth and first order the Eq.~(\ref{sigmaeq}) is an elliptic PDE for $N$ that fixes it for any time. This implies that at zeroth and first order in $L^{-1}$ the  equation of motion (\ref{dotn}) must not be regarded as an independent evolution equation since the evolution of $N$ is derived from the other canonical variables. Equation (\ref{dotn}) is instead the equation that determines $\tau$ at these orders. Therefore, the evolution of the system at zeroth and first order in $L^{-1}$ does not require $N(t=0,\vec{x})$ as an independent initial data. It is involved in the initial value problem from second order in $L^{-1}$ and higher. 


\section{Conclusions}
We have carried out the Hamiltonian analysis for a Ho\v{r}ava model with a potential composed of $R$- and $R^2$-terms. We have performed Dirac's algorithm for extracting all the constraint of the theory and ensuring their preservation in time, showing in this way the closure of the algebra of constraints. 

The constraint emerging when the preservation in time of the Hamiltonian is imposed can be solved for the variable $\pi = g_{ij} \pi^{ij}$ in a closed way, since this constraint is an elliptic PDE for $\pi$. Then, the preservation in time of this constraint is imposed and it turns out that Dirac's algorithm ends at this step since this yields a first-order PDE for a Lagrange multiplier of the theory. This equation is a flow equation along the integral lines of a divergenceless vector field and its solutions are compatible with the expected asymptotic behavior of all the fields. Our analysis provides non trivial evidence about the consistent structure of constraints for Ho\v{r}ava models that go beyond the lowest-order potential.

We have seen that the inclusion of the $R^2$ term in the potential gives rise to an additional scalar degree of freedom that is absent in the $\lambda R$-model analyzed in Ref.~\cite{Bellorin:2010je}. This scalar is an odd mode in the sense that it propagates only one field as Cauchy data. This agrees with the extra mode found in Refs.~\cite{Li:2009bg,Blas:2009yd}. We have also performed a linear-order perturbative analysis on a weakly varying background, showing the decoupling of the extra mode for zeroth- and first-derivatives of the background. We have discussed how the initial data associated to the extra mode smoothly decouples from the initial value problem of the whole system.

The physical propagation of the odd extra mode is an analysis that must be done carefully, due to its peculiarity of having an evolution equation of first order in time-derivative, as was also noticed in Ref.~\cite{Blas:2009yd}. In any case, the analysis should effectively include higher-order terms in the potential.

Our results indicate the classical consistency of the truncated $\lambda R^2$-model. However, the validity of the full nonprojectable Ho\v{r}ava theory still remains unanswered. In fact, as we have mentioned, even in perturbation theory the propagation of the extra mode has not been completely analyzed. Moreover, the quantization of the Ho\v{r}ava theory is a challenging task due to the presence of very involved second-class constraints.


\appendix
\section*{Appendix: Analysis of the equation for $\sigma$}
\setcounter{section}{1}
The main difference between the Eq.~(\ref{sigmaeq}) for $\sigma$ and the standard constraints of general relativity is that Eq.~(\ref{sigmaeq}) is not an elliptic PDE. This deserves some discussion about the nature of the differential operator of Eq.~(\ref{sigmaeq}) and the boundary data compatible with it.

We consider the existence and uniqueness of the solutions of Eq.~(\ref{sigmaeq}) on a $t = \mbox{constant}$ slice $\mathcal{M}_3$, which is assumed to be a Riemannian manifold topologically equivalent to $\mathbb{R}^3$. We may rewrite (\ref{sigmaeq}) as 
\begin{equation}
 V^i \partial_i (\sigma / N) = \mathcal{J} \,,
\label{sigmaeq2}
\end{equation}
where the source $\mathcal{J}$ is independent of $\sigma$. All fields are assumed to be smooth geometrical objects on $\mathcal{M}_3$. 

The l.h.s.~of Eq.~(\ref{sigmaeq2}) is the application of the divergenceless (constraint $\mathcal{C}$) vector field $V$, whose components are $V^i = N^2 \gamma^i /\sqrt{g}$, on the scalar field $\sigma / N$. Therefore, in order to analyze the Eq.~(\ref{sigmaeq2}) it is important to describe the flow generated by $V$. To this end we shall consider some mild assumptions on $V$, which in particular may be imposed on the initial data over the slice $t = 0$ when formulating the initial value problem to the field equations of the full theory: We assume that the zeros of $V$ belong to a 2D surface $\Sigma \subset \mathcal{M}_3$, where $\Sigma$ splits $\mathcal{M}_3$ into two open sets denoted by $\mathcal{M}_+$ and $\mathcal{M}_-$. $\Sigma$ may be extended to infinity or it may be a compact smooth surface without boundaries.

The flow determined by the vector field $V$ is obtained as the solution of the ODEs
\begin{equation}
 \frac{dx^i}{ds} = V^i (x^. (s)) \,,
 \label{dx}
\end{equation}
where $x^i$ are local coordinates on $\mathcal{M}_3$. The solution $ x^i = x^i (s,s_1,s_2)$ depends on the flow parameter $s$ and $s_1,s_2$, where $x^i(0,s_2,s_2)$ is the local parametric description of $\Sigma$. In turn, the Eq.~(\ref{sigmaeq2}) can be cast as
\begin{equation}
 \frac{d(\sigma/N)}{ds} =
 \mathcal{J}(x^.(s,s_1,s_2)) \,.
 \label{dsigma}
\end{equation}
The system of ODEs (\ref{dx}) - (\ref{dsigma}) is equivalent to the PDE (\ref{sigmaeq2}) in a neighborhood of $\Sigma$. The change of variables $ x^i \leftrightarrow (s,s_1,s_2)$ is a diffeomorphism in this neighborhood.

Let us denote by $u$ the unit vector field orthogonal to $\Sigma$ and oriented inwardly $\mathcal{M}_+$. Let
\begin{equation}
 U_{\pm} =
 \left\{ p \in \Sigma \; : \; V \cdot u \gtrless 0 \right\}
\hspace{2em}
\mbox{and}
\hspace{2em}
 U_0 = \left\{ p \in \Sigma \; : \; V \cdot u = 0 \right\} \,.
\end{equation}
The continuity assumption ensures that any curve on $\Sigma$ from $p_1 \in U_+$ to $p_2 \in U_-$ necessarily intersects $U_0$. Notice that $U_+$ by definition is automatically a noncharacteristic surface with respect to the vector field $V$. Our aim is to show that $U_+$ is the appropriated surface to give the boundary condition on $\sigma / N$ in order to solve uniquely the Eq.~(\ref{sigmaeq2}) on $\mathcal{M}_+$.

\begin{figure}[t]
 \includegraphics[scale=0.3]{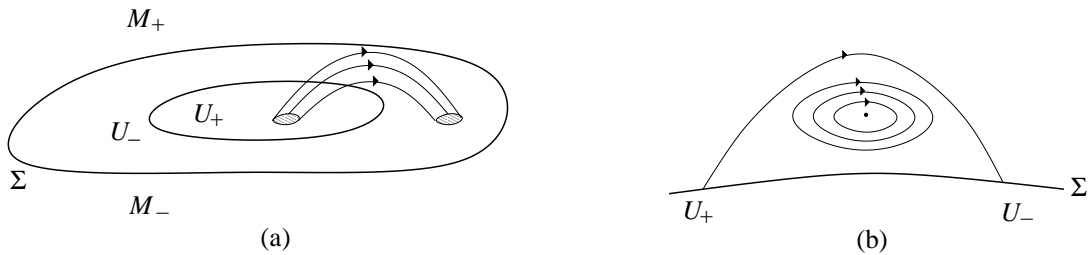}
 \caption{(a) A typical flow of $V$. (b) If a zero of $V$ was excluded from $\Sigma$ then in the neighborhood of the zero there could be a set of loops disconnected to $\Sigma$.}
 \label{fig:typicalflow}
\end{figure}

There are two important facts resulting as consequences of the constraint on $V$ and the settings we are considering. The first one concerns to what extent the flow of $V$ can reach the infinity. We remark that $V$ is divergenceless and its asymptotic behavior ($\mathcal{O}(r^{-3})$) ensures that its flux at infinity is always zero. Thus, one can take a Gaussian surface in $\mathcal{M}_+$ with the form of a tube parallel to the flow lines of $V$ and having one of its caps coinciding with an arbitrary subset $\Omega \subset U_+$ of nonzero measure and the other cap at infinity. By considering the flux of $V$ on the whole Gaussian surface one obtains automatically that the flux on $\Omega$ vanishes. Therefore, it turns out that if we take points on $U_+$, only a set of measure zero, if any, may have associated curves $x^i(s,\hat{s}_1,\hat{s}_2)$ ending at infinity. The main flow of $V$ always ends at $U_-$ or $U_0$ [see Fig.~\ref{fig:typicalflow}.a].

The second point is the fact that the flow of $V$ starting on points of $U_+$ defines a map $x^i (s,s_1,s_2): U_+ \rightarrow \mathcal{M}_+$ which is a surjective map on $\mathcal{M}_+$. In this point we remark that, since $\Sigma$ contains all the zeros of $V$, if a loop on $\mathcal{M}_3$ circulates around a zero we avoid situations like the one showed in Fig.~\ref{fig:typicalflow} (b). In that case the flow starting at $U_+$ does not span completely $\mathcal{M}_+$, hence the flow does not determine $\sigma / N$ inside the loop. In our case any loop on $\mathcal{M}_3$ circulating a zero necessarily intersects $\Sigma$.

After these considerations we have arrived to our main conclusion about the equation for the Lagrange multiplier $\sigma$. The standard theorems on ODEs ensure the local existence and uniqueness of the system (\ref{dx}) - (\ref{dsigma}) and hence of the PDE (\ref{sigmaeq2}) on $\mathcal{M}_+$: given an initial data $f_+$ on any open set $\Omega \in U_+$, there always exists a unique smooth solution $\sigma / N$ of Eq.~(\ref{sigmaeq2}) in a neighborhood $\Omega$ satisfying $( \sigma / N ) |_{\Omega} = f_+$.

Our qualitative analysis suggests that the solution may be extended globally. However, in order to address that problem a rigorous discussion of the propagation around the zeros of the vector field $V$ must be considered. We will analyze this problem elsewhere.

The Dirac algorithm for the $\lambda R^2$-model ends at this step since the Eq.~(\ref{sigmaeq}) is solved for the Lagrange multiplier $\sigma$. However, it is interesting to notice that $\sigma$ is not completely determined by Eq.~(\ref{sigmaeq}): the initial data $f_+$ at $U_+$ is arbitrary and the solution $\sigma$ depends continuously on it. This means that the associated primary constraint $\phi$ is a mixture of first- and second-class constraints. The gauge symmetry generated by the first-class part has a gauge parameter depending only on the local coordinates on $\Sigma$. In particular, this lower-dimensional gauge symmetry may be used to fix a boundary condition on $\Sigma$ for the scalar field $N$. Besides this subtlety, $N$ describes a half-degree of freedom that is absent in general relativity.


\end{document}